\documentclass[]{emulateapj}
\usepackage{graphicx}
\usepackage{hhline}
\usepackage{xcolor}

\usepackage[normalem]{ulem}

\usepackage{amssymb, amsmath}

\newcommand{\Ms}{M_{\odot}}
\newcommand{\Mb}{M_{\rm BH}}
\newcommand{\Mz}{M_{\rm ZAMS}}
\newcommand{\Mnb}{M_{\rm NS, b}}
\newcommand{\Mng}{M_{\rm NS, g}}

\newcommand{\fe}{f_{\rm ej}}

\begin{document}

\title{Confronting Models of Massive Star Evolution and Explosions with Remnant Mass Measurements}
\author{Carolyn A. Raithel\altaffilmark{1}, Tuguldur Sukhbold\altaffilmark{2}$^{,}$\altaffilmark{3}, Feryal \"Ozel\altaffilmark{1}$^{,}$\altaffilmark{4}}

\altaffiltext{1}{Department of Astronomy and Steward Observatory, University of Arizona, 933 N. Cherry Avenue, Tucson, Arizona 85721, USA}
\altaffiltext{2}{Department of Astronomy, The Ohio State University, Columbus, 140 W. 18th Ave., Columbus, OH 43210, USA}
\altaffiltext{3}{Center for Cosmology and AstroParticle Physics, The Ohio State University, 191 W. Woodruff Ave., Columbus, OH 43210, USA}
\altaffiltext{4}{Guggenheim Fellow}

\begin{abstract}
The mass distribution of compact objects provides a fossil record that can be studied to uncover information on the late stages of massive star evolution, the supernova explosion mechanism, and the dense matter equation of state. Observations of neutron star masses indicate a bimodal Gaussian distribution, while the observed black hole mass distribution decays exponentially for stellar-mass black holes. We use these observed distributions to directly confront the predictions of stellar evolution models and the neutrino-driven supernova simulations of \citet{Sukhbold2016}. We find excellent agreement between the black hole and low-mass neutron star distributions created by these simulations and the observations. We show that a large fraction of the stellar envelope must be ejected, either during the formation of stellar-mass black holes or prior to the implosion through tidal stripping due to a binary companion, in order to reproduce the observed black hole mass distribution. We also determine the origins of the bimodal peaks of the neutron star mass distribution, finding that the low-mass peak (centered at $\sim1.4~\Ms$) originates from progenitors with $\Mz\approx9-18~\Ms$. The simulations fail to reproduce the observed peak of high-mass neutron stars (centered at $\sim1.8~\Ms$) and we explore several possible explanations. We argue that the close agreement between the observed and predicted black hole and low-mass neutron star mass distributions provides new promising evidence that these stellar evolution and explosion models are accurately capturing the relevant stellar, nuclear, and explosion physics involved in the formation of compact objects.
\end{abstract}

\maketitle

\section{Introduction}

The masses of compact objects that are formed through massive star evolution are relics of the various physical processes that take place during the star's lifetime and subsequent death. First, whether the star explodes and forms a neutron star or implodes to form a black hole is largely determined by the advanced-stage evolution in the stellar core. Beyond the star's fate, the pre-supernova core structure also influences the mass of the resulting neutron star following a successful explosion. If instead the star implodes, the black hole mass is affected by the star's mass loss history. Second, the key processes that take place during the core collapse itself, such as neutrino transport and multi-dimensional turbulence, can also influence the nature of the outcome. Furthermore, the dense matter equation of state plays a role in determining the possible range of neutron star and black hole masses, setting both the maximum neutron star mass and potentially the minimum black hole mass. Because the mass distribution of compact objects is collectively shaped by each of these processes, it has the potential to provide insight into the fundamental physics underlying stellar evolution, the supernova (SN) explosion mechanism, and the dense matter equation of state.

Observationally, the mass distribution can be inferred from the known sample of neutron star and black hole masses. Many black hole masses have been measured from X-ray binary systems, while over 30 precision neutron star masses have been measured from double neutron stars and millisecond pulsars (for a recent review of the latter, see \citealt{Ozel2016}).  To infer the black hole mass distribution, \citet{Ozel2010b} combined measurements from 16 low-mass X-ray binaries, finding that it follows an exponential decline. \citet{Farr2011} fit a similar sample of black hole masses from 15 low-mass X-ray binaries, but also included black holes from 5 high-mass X-ray binaries. They found that the low-mass population follows a power-law distribution, while the combined population follows an exponential decline. The mass distribution of neutron stars has also been measured, with an ever-growing and precise sample \citep{Thorsett1999, Ozel2012, Kiziltan2013, Antoniadis2016}. The most recent study by \citet{Antoniadis2016} inferred a bimodal distribution, possibly indicating two separate formation channels for creating neutron stars.

From the theoretical side, there have been recent new developments in our understanding of the evolution of massive stars and the modeling of their explosions. In particular, during the advanced stages of core evolution of massive stars, the interplay of convective burning episodes of carbon and oxygen gives rise to final pre-supernova structures that are non-monotonic in  initial mass \citep{Sukhbold2014}. The pre-supernova core structure of a massive star, i.e., the density gradient surrounding the iron core, is known to play a pivotal role in determining whether the star explodes or implodes (e.g., \citealt{Burrows1995}). Several recent studies have explored the connection between this final core structure and the landscape of neutrino-driven explosions of massive stars through numerical and semi-analytical approaches \citep{OConnor2011, Ugliano2012, Pejcha2015, Ertl2016, Sukhbold2016, Muller2016, Murphy2017}. While these works differ in their scope and complexity, all find that there is no single initial mass below which stars only explode and above which only implode. Instead, they recover a much more complicated landscape of explosions as dictated by the pre-supernova evolution of massive stars.

A number of previous studies have explored the connection between the supernova mechanism and the observed distribution in compact objects. For example, \citet{Pejcha2012} used the mass distribution of double neutron stars to constrain the entropy coordinate in the progenitor at which the explosion must originate. In another work, \citet{Kochanek2014} related the observed mass distribution of black holes to the core compactness of the progenitor star, in order to constrain core-collapse SN models. Both studies, however, used mass cuts, rather than realistic explosion simulations, to determine the predicted remnant masses. 

Sukhbold et al. (2016; hereafter, S16) surveyed the explosion outcomes, including the nucleosynthesis yields, light curves, and compact remnants, for a large set of stellar models using a novel one-dimensional neutrino-driven explosion mechanism, based on \citet{Ugliano2012} and \citet{Ertl2016}. While three-dimensional models are the gold standard in understanding SN explosions, they are computationally expensive and prohibit an exploration of a large parameter space. Furthermore, although great progress is being made in multi-dimensional explorations of the problem, a consensus has not yet been reached in the community (\citealt{Janka2016}, and references therein). Though simplified, the one-dimensional treatment in S16 allows for large parameter-space studies. In that work, the authors explored the outcomes due to various parameterizations of the central engine applied to 200 pre-SN stars with initial masses between 9 and 120~$\Ms$. They performed a preliminary comparison of the remnant masses produced in their simulations to the observed populations of black holes and neutron stars, and found reasonable agreement in the produced mass range. However, the comparison was qualitative as they did not properly weight the observed masses or explore any observational biases. 

This new, fine grid of stellar evolution models, combined with a simplified, parametric explosion mechanism and combined with a better quantitative understanding of remnant mass distributions, now make it possible to confront these theoretical models with remnant mass observations in a systematic way. One potential difficulty in such a comparison, however, is that the theoretical models describe single-star evolution, while all precision mass measurements come from binary systems. We argue that a meaningful comparison can nevertheless be made for the following reasons. 

For late-time mass transfers (cases B and C; \citealt{Smith2014}), the He-core mass, which is the main determinant for the final pre-SN structure \citep{Sukhbold2014}, is not appreciably affected by the mass transfer. Thus, for these scenarios, the remnant outcome will not be strongly affected by  binary evolution. On the other hand, in early stable mass transfers via Roche-lobe overflow, the He core mass can be affected. However, this effect can be at least partially encompassed by a stronger mass loss efficiency in the single-star models. In other words, even though the models here describe single star evolution, due to the uncertain nature of mass loss \citep[e.g.,][]{Renzo2017}, the single star models can be representative of some close binary scenarios. We revisit this point in $\S$\ref{sec:binary}.
For millisecond pulsars, which are spun up by accretion from their binary companion \textit{after} forming, \citet{Antoniadis2016} found that the accretion rates onto the neutron star are very inefficient, and that the observed masses must be close to their birth masses. Thus, we take the approach in this paper that comparing the remnant masses measured from binary systems to theoretical models of single-star evolution indeed can provide a reasonable first constraint on the theoretical models.

With these motivations, we directly confront the stellar evolution models and SN outcomes of S16 with the observed black hole and neutron star mass distributions. We describe the stellar evolution and supernova models in more detail in $\S$\ref{sec:models}. In $\S$\ref{sec:obs}, we review the current collection of observed compact object masses. In $\S$\ref{sec:BHdist}, we compute the simulated black hole mass distribution, in order to compare it to the observed distribution on a level playing field. We find that the fraction of the stellar envelope that must be ejected during the SN implosion in order to recreate the observed black hole mass distribution is quite large. In $\S$\ref{sec:NSdist}, we compute the mass distribution of neutron star remnants and find surprisingly close agreement between the simulated outcomes and low-mass peak of the observed bimodal distribution of \citet{Antoniadis2016}. We explore the origin of this peak and find that it originates from progenitors with zero-age main sequence masses ($\Mz$) in the range $\Mz \approx 9-18~\Ms$. We discuss in $\S$\ref{sec:highM} possible explanations for the lack of high-mass neutron stars and LIGO-mass black holes in the simulations. Finally, we discuss the possible implications of our inferred distributions for stellar evolution and explosion models more generally.

\section{Pre-SN Stellar Evolution and Explosion}
\label{sec:models}
The nucleosynthesis yields, remnant masses, and light curves due to neutrino-driven explosions from a wide range of solar metallicity massive stars were surveyed recently in S16. In the following, we briefly highlight aspects of that work that are relevant to the present study.

All of the progenitor models used in S16 were computed using the one-dimensional implicit hydrodynamics code \texttt{KEPLER} \citep{Weaver1978}. The main progenitor set consists of 200 non-rotating, solar metallicity models with initial masses between 9 and 120~$\Ms$, and were mostly compiled from \citet{Woosley2007}, \citet{Sukhbold2014}, and \citet{Woosley2015}.  Between the initial masses of 13 and 30~$\Ms$, the models were calculated with 0.1~$\Ms$ increments. As will be shown in $\S$\ref{sec:NSdist}, the large number of models with fine resolution in initial mass space were critical in uncovering discrete branches in the neutron star mass distribution.

Mass loss rates from \citet{Nieuwenhuijzen1990} were employed in all models. While the lightest stars don't lose much mass throughout their evolution, the mass loss gradually strengthens with increasing initial mass. The entire envelope was lost for stars with initial masses above 40~$\Ms$ and a Wolf-Rayet wind from \citet{Wellstein1999} was adopted for these stars. Due to coarse sampling in mass space for high-$\Mz$ stars and due to the assumed input physics, nearly all Wolf-Rayet pre-SN stars lost their He-envelopes as well, and, therefore, died as carbon-oxygen (CO) cores.

Although the He core mass, and hence the final pre-supernova structure, is insensitive to mass loss for the lighter stars that don't lose all of their envelope, the masses of black holes that are formed if the star implodes have an appreciable dependence on the adopted prescription of mass loss. For stars with $\Mz < 40~\Ms$, the amount of envelope remaining sets the range of possible black hole masses upon implosion, while for stars with $\Mz>40~\Ms$, the final star mass approximately sets the possible black hole mass.

The final fates of \texttt{KEPLER} pre-supernova progenitors were modeled from the onset of core collapse through core bounce and to either a successful explosion or implosion with the Prometheus-Hot Bubble code (\texttt{P-HOTB}, \citealt{Janka1996, Kifonidis2003}). \texttt{P-HOTB} is a one-dimensional Eulerian hydrodynamics code with a simplified gray neutrino-transport solver and a high density equation of state (\citealt{Lattimer1991}, with K=220 MeV). The simulations were run for sufficiently long times in order to determine the final mass cuts and explosion energies. For technical details and further discussion see \citet{Ugliano2012} and \citet{Ertl2016}.

A major improvement of the recent studies such as S16, \citet{Ugliano2012}, and \citet{Ertl2016} over the previous surveys is that the explosion outcomes were free from arbitrary mass cuts and directly dialed-in explosion energies (e.g., \citealt{Woosley2002, Chieffi2013, Nomoto2013}). This was achieved by calibrating the free parameters of an analytic proto-neutron star (PNS) cooling model to reproduce the observables of SN1987A for five different models of the progenitor star: W15, W18, W20, N20, and S19.8. Each model resulted in a particular choice of parameters (which we call the ``central engine," henceforth) and each central engine was then applied to the 200 \texttt{KEPLER} pre-supernova models. Furthermore, S16 improved the low ZAMS-mass end compared to \citet{Ugliano2012} by adding SN~1054 as a calibration anchor and interpolating the core parameters to account for the reduced PNS contraction in the case of small PNS masses (see S16 for details). Finally, each explosion yielded a unique set of observational outcomes, including the remnant mass, that is characterized by the pre-supernova core structure of the progenitor. This, in turn, allowed us to construct the expected compact remnant mass distributions for each central engine.

\begin{figure}[ht]
\centering
\includegraphics[width=0.48\textwidth]{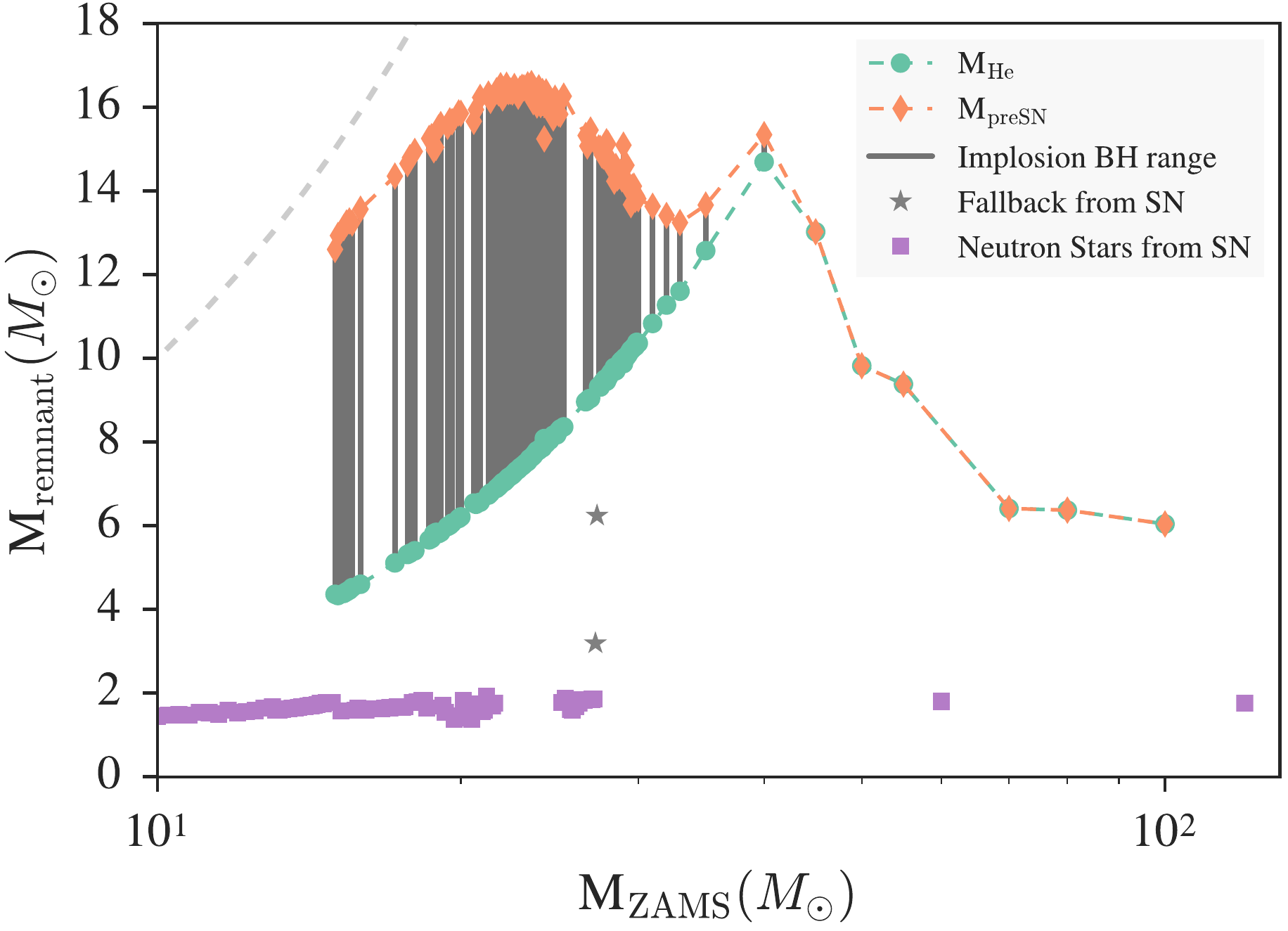}
\caption{\label{fig:allrem} Baryonic remnant masses as a function of the progenitor ZAMS mass, for the central engine W18. Neutron star remnant masses from successful explosions are shown in purple. The range of possible black hole masses, shown in gray, is bound by the He-core mass (green circles) and pre-SN mass (orange diamonds) of the progenitor, because an uncertain fraction of the stellar envelope may be ejected either during the formation of the black hole, via the Nadyozhin-Lovegrove mechanism \citep{Nadezhin1980, Lovegrove2013}, or prior to the implosion by tidal stripping from a binary companion. The gray dashed line denotes the initial progenitor mass. Note the co-existence of neutron star and black hole outcomes between $\Mz \sim15-21~\Ms$ and 25$-$28$\Ms$.}
\end{figure}

Figure~\ref{fig:allrem} shows the baryonic remnant masses produced by one sample engine (W18) as a function of the initial progenitor mass. A successful explosion results in a neutron star for most models (purple), but in a very small number of cases that experienced significant amount of fallback, a light black hole is formed (gray stars). Since such fallback black holes occur infrequently and only at relatively high mass models, we omitted them from our analysis.\footnote[1]{We verified that the fallback black holes do not affect our conclusions by repeating the analysis described in $\S$\ref{sec:BHdist} and including an additional branch to model them with a Gaussian distribution. We found no significant change to our results.}

Although a failed explosion would certainly form a stellar-mass black hole, its exact mass is not well determined for progenitors that retain some amount of envelope by the time of implosion (i.e., for stars with $\Mz < 40~\Ms$). A weak shock, which is launched by the loss of the proto-neutron star binding energy in the Nadyozhin-Lovegrove effect, may be able to eject a fraction or all of the remaining envelope \citep{Nadezhin1980, Lovegrove2013, Coughlin2017, Fernandez2017}. Additionally, it is possible that some of the progenitor envelope may be stripped by a binary companion prior to the implosion. If common Type I-b/c SNe arise from progenitors that have lost their envelope to a companion (e.g., \citealt{Dessart2012, Dessart2015}), it would not be surprising to if some of these stripped cores fail to explode. Thus, we might expect that a fraction of all remnant black holes come from the collapse of stripped cores. The black hole masses from stellar implosions (gray lines in Fig.~\ref{fig:allrem}) are thus bounded by the He-core (green circles) and the final pre-SN (orange diamonds) masses of the progenitor, and ultimately depend on how much of the stellar envelope gets ejected during or prior to the implosion. We analyze this further in $\S$\ref{sec:BHdist}.

\begin{figure}[ht]
\centering
\includegraphics[width=0.48\textwidth]{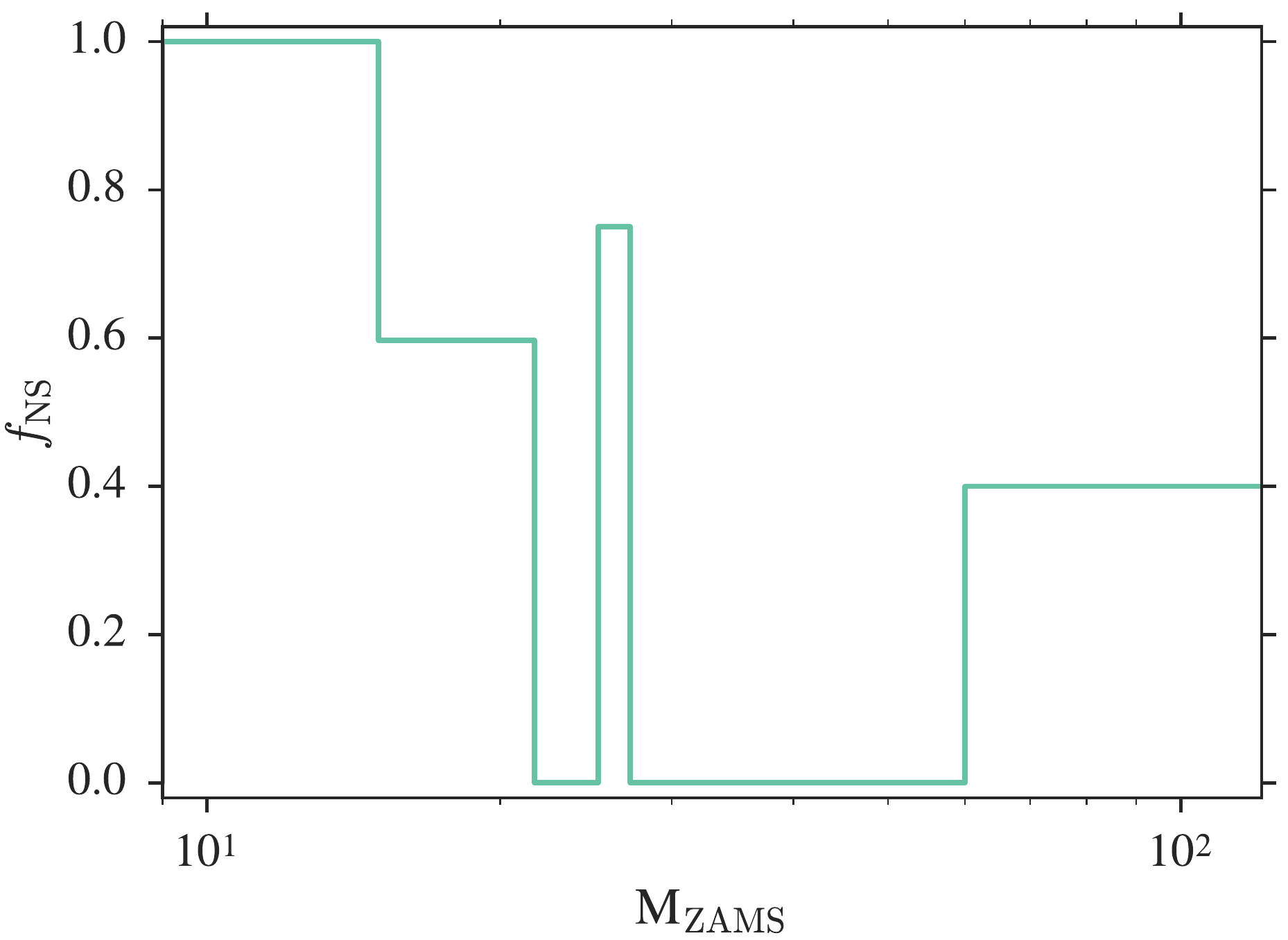}
\caption{\label{fig:fNS} Fraction of neutron stars formed as a function of progenitor initial mass, for the outcomes shown in Fig.~\ref{fig:allrem}. The initial mass has been binned to produce the minimum number of bins while still capturing whether a mass region has only neutron star outcomes, only black hole outcomes, or some combination of the two.}
\end{figure}

Finally, we note the presence of large intervals in $\Mz$-space over which the outcomes switch between neutron stars and black holes in Fig.~\ref{fig:allrem}. As has previously been reported, the explodability of the pre-SN star is not determined by only the initial mass \citep{OConnor2011, Ugliano2012, Pejcha2015, Ertl2016, Sukhbold2016, Muller2016, Murphy2017}. For example, \citet{Ertl2016} propose a new two-dimensional parameter space to characterize pre-SN stars and to predict whether a neutron star or a black hole forms following core collapse. Specifically, a separatrix can be drawn in this space such that any model that falls above it will explode, while models below it will implode. In this framework, we interpret the large intervals of neutron star or black hole outcomes to result from repeated crossings of this separatrix as the initial mass varies and the core structure changes. In addition, there are also smaller intervals in initial-mass space, such as between $\Mz \sim 15$ and 21~$\Ms$ and between 25 and 28~$\Ms$ in which the locus of pre-SN stars in this parameter space straddles the separatrix and, correspondingly, the outcome frequently changes. As a result, even very small changes in the initial conditions may turn a successful explosion into an implosion or vice versa. We, therefore, interpret the outcomes in these mass ranges not as rapid oscillations between the two types of remnants but rather as the co-existence of two channels with different likelihoods (see also the discussion in \citealt{Clausen2015}). We show this probabilistic interpretation of outcomes in Fig.~\ref{fig:fNS}, where we plot the relative fraction of neutron stars produced, $f_{\rm NS}$, as a function of the initial mass. We identify several regions where the outcomes can be only neutron stars,  only black holes, or a combination of the two. We use these branches to appropriately weight the remnant outcomes when comparing the simulated and observed mass distributions in $\S$\ref{sec:BHdist}-\ref{sec:NSdist}.

\section{Observations of Remnant Masses}
\label{sec:obs}
\begin{figure}[ht]
\centering
\includegraphics[width = 0.45 \textwidth]{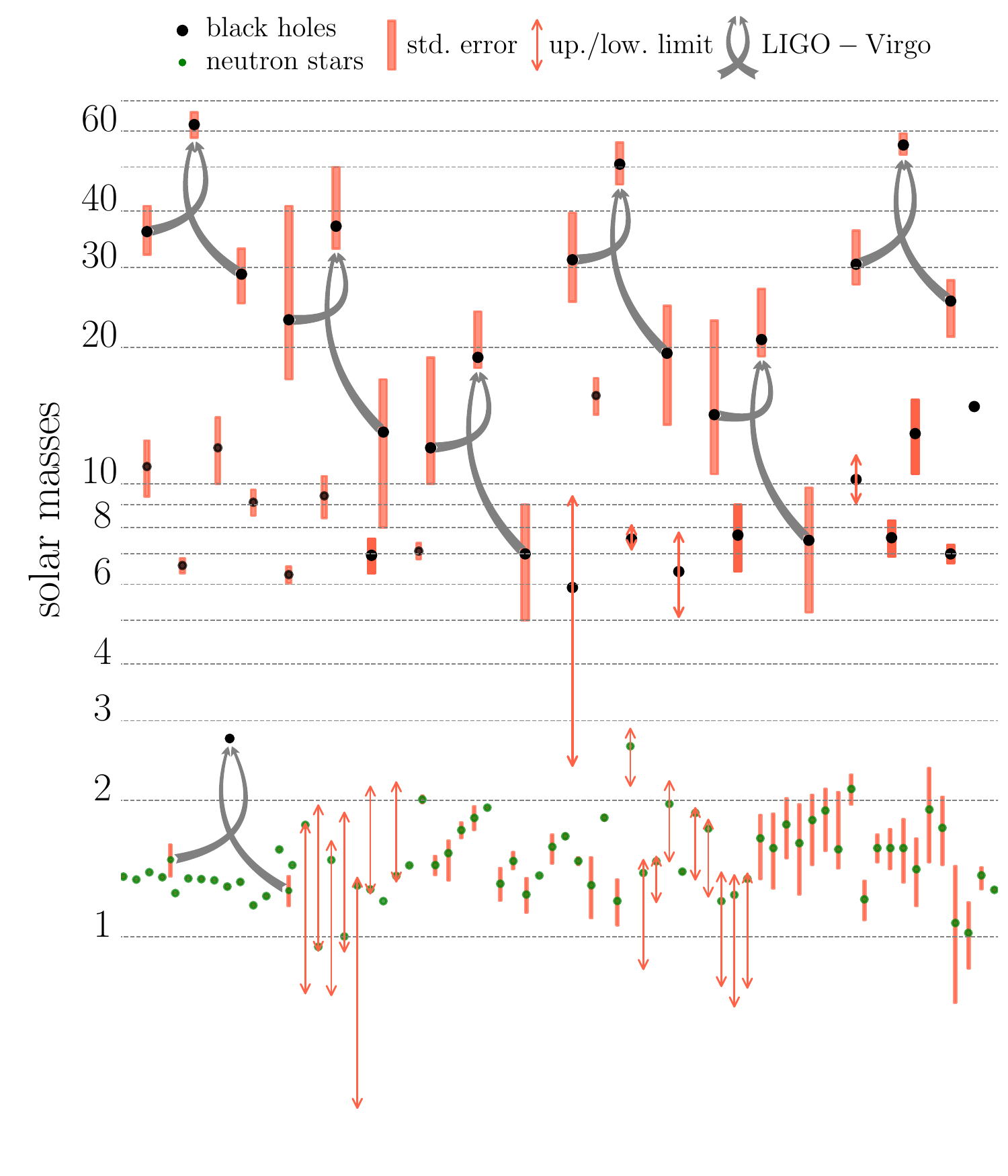}
\caption{\label{fig:graveyard}  Observed masses of neutron stars and black holes. The green points show neutron stars, while the black points show black holes. The red vertical lines represent the error bars for each measurement. The red vertical arrows denote upper and lower measurement limits, and should not be taken as Gaussian uncertainties. The gray arrows connect the progenitors to the outcome mass for each of the six confirmed and one candidate detection of gravitational waves from merging black holes and neutron stars.}
\end{figure}

The simulation outcomes described in $\S$\ref{sec:models} can be directly compared to the observed masses of compact objects, as we will describe in $\S$\ref{sec:BHdist} and \ref{sec:NSdist}. First, however, we review the current status of the measurements. The observed masses, which are summarized in Fig.~\ref{fig:graveyard}, come from a few primary types of observations:  timing and spectra of X-ray binaries containing stellar-mass black holes, timing of millisecond pulsars with white dwarf companions as well as of the double neutron stars, and, most recently, the detection of gravitational waves from black hole-black hole mergers and the first neutron star-neutron star merger.

Black hole masses are typically measured dynamically from X-ray binaries. Data on 23 confirmed black hole X-ray binaries and on 32 additional transient systems with candidate black hole members have been compiled in \citet{Ozel2010b}. From these data, masses are provided for 16 confirmed black holes, based on some combination of constraints on the mass ratio, inclination angle, or the mass function for the system.
A similar compilation can be found in \citet{Farr2011}, which focused on the masses of 15 black holes in low-mass X-ray binaries undergoing Roche-lobe overflow, as well as 5 black holes in wind-fed, high-mass X-ray binaries. Several masses in Fig.~\ref{fig:graveyard} are also taken from the more recent compilation found on the StellarCollapse website,\footnote{http://www.stellarcollapse.org/bhmasses} from \citet{Wiktorowicz2014}, and the references therein. Finally, the most recent estimate on GX 339-4 by \citet{Heida2017} has also been included in Fig.~\ref{fig:graveyard}.

Neutron star masses are also measured dynamically, but with different methods. While spectra from an optically bright companion can be used to constrain the neutron star mass, the majority of masses are measured from radio pulsar timing. In particular, for millisecond pulsars with white dwarf companions, the measurement of any post-Keplerian parameters in the pulsar timing residuals can be used to constrain the pulsar mass, when combined with the mass function. Precision masses for 32 millisecond pulsars were recently summarized in \citet{Antoniadis2016}. Precision masses for the sub-population of double neutron stars are also determined from the timing measurement of at least two post-Keplerian parameters. For a recent review of all neutron star mass measurements, see \citet{Ozel2016}.\footnote{The neutron star masses can be found at http://xtreme.as.arizona.edu/NeutronStars/data/pulsar\_masses.dat}

Finally, there has been a new addition to these families of mass measurements, thanks to the first detections of gravitational waves by LIGO and subsequently, the LIGO-Virgo Collaboration. To date, five confirmed detections have been made from the mergers of binary black holes \citep{Abbott2016b,Abbott2016,Abbott2017,GW170608, LIGOVirgo2017}. The sixth set of gravitational wave black holes shown in Fig.~\ref{fig:graveyard} is from the candidate black hole-black hole merger LVT151012 \citep{Abbott2016c}. As shown in Fig.~\ref{fig:graveyard} and as we will discuss more in $\S$\ref{sec:LIGO}, some of these black hole masses are larger than had previously been observed or even thought possible. Additionally, the first detection of gravitational waves from a neutron star-neutron star merger was recently announced \citep{Abbott2017a}, and offers a new way of adding measurements to the collection of neutron star masses as well.

The most uncertain of these masses are those that come from the measurement of a single post-Keplerian parameter in a neutron star or black hole binary, with no additional constraints on the system. Such measurements provide only an upper and lower limit to the component masses, which we represent in Fig.~\ref{fig:graveyard} with vertical red arrows. These arrows represent a likely mass range for an assumed isotropic distribution of binary inclination and should not be interpreted as Gaussian uncertainties. We also note that Fig.~\ref{fig:graveyard} does not include any measurements with only an upper or lower limit; we include only points with both an upper and lower limit, or with well-defined error bars. All error bars (shown as the solid red lines) represent the 68\% confidence intervals, except for the LIGO masses, which denote 90\% confidence intervals. Finally, we note that the black hole mass measurement for NGC~300~X-1 has been excluded due to the likely asymmetric irradiation of stellar winds, which contaminates the mass measurement (Tom Maccarone, priv. communication).

\section{Black hole mass distribution}
\label{sec:BHdist}
\begin{figure*}[ht]
\centering
\includegraphics[width = 0.9 \textwidth]{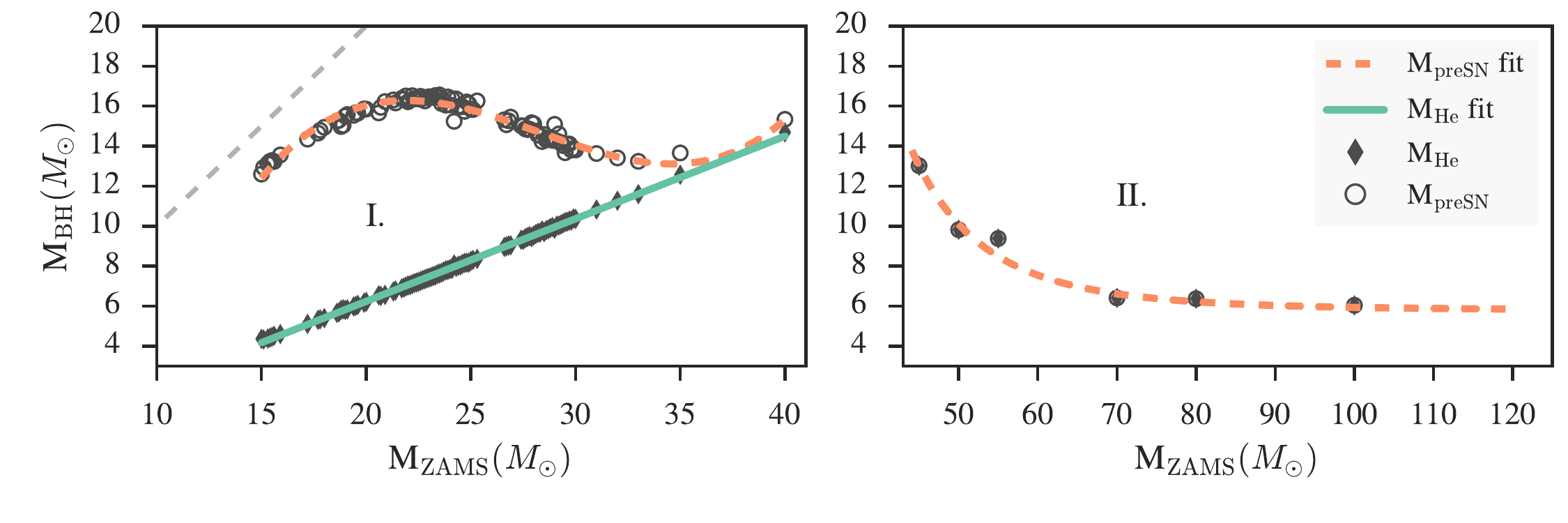}
\caption{\label{fig:BHrem} Same as Fig.~\ref{fig:allrem}, but only showing the implosion black holes. Solid diamonds indicate black holes that are formed from the collapse of only the He- or CO-core; open circles indicate black holes formed from the collapse of the entire star. We also label the two progenitor mass ranges, across which we fit the black hole masses with analytic functions. We show the fits to the remnants of implosions in which the entire star collapsed and in which only the core collapsed in orange dashed and green solid lines, respectively. }
\end{figure*}

Our goal is to directly compare the outcomes of the S16 stellar evolution and SN simulations with the measured remnant masses discussed above. We first focus on the models that produce remnant masses larger than the maximum, theoretically-allowed neutron star mass and, therefore, yield black holes.

The observed masses of compact objects have previously been fit with simple functional forms. The functional forms have been chosen to provide a theoretically motivated description of the data and to help facilitate a direct comparison between the observations and theory. We, therefore, start by creating analytic functions to model the remnant masses as a function of their ZAMS progenitor masses. By convolving these functions with the initial mass function (IMF), we determine the expected distribution, which can then be directly compared to the observed distribution.

Figure~\ref{fig:BHrem} shows the black hole remnant masses calculated in the simulations described in $\S$\ref{sec:models} for engine W18. For simplicity, we only show the outcomes from one particular engine in Fig.~\ref{fig:BHrem}, but we include results from all of the following five engines from S16 in our analysis:  W15, W18, W20, N20, and S19.8. Each model produces qualitatively similar results to those shown in Fig.~\ref{fig:BHrem}, so we average the results from each model in the following analysis.

In Fig.~\ref{fig:BHrem}, we also identify two different branches of mass outcomes, so that each branch is well-approximated by a single function. The branches are divided as follows: Branch I spans $15 \le \Mz \le 40 \Ms$, which corresponds to the range of red supergiant models that retained significant amounts of the envelope upon implosion. Branch II spans $45 \le \Mz \le 120 \Ms$, which corresponds to the range of models that lost all of their envelopes and die as Wolf-Rayet stars.

In Branch I of the black hole masses, the outcomes are bounded by two possibilities: (1) He-core implosion, which occurs in the event that the entire hydrogen envelope is ejected by a weak shock during the black hole formation or has been tidally stripped by a binary companion prior to the collapse, or (2) implosion of the entire pre-SN stellar mass. In our modeling of these outcomes, we allow for a variable fraction of the envelope to be ejected, which we denote as $\fe$. For the more massive progenitors in Branch II, which do not retain their hydrogen and helium envelopes, $\fe$ has no physical meaning. For these models, the only scenario considered is the collapse of the CO-core.

Accordingly, the filled diamonds in Fig.~\ref{fig:BHrem} indicate that the entire stellar envelope was ejected prior to or during the implosion and only the core collapsed to form the black hole. We represent remnant masses from this scenario as M$_{\rm BH,core}$, which have an ejection fraction, $\fe=1$. On the other end of the spectrum, open circles indicate that the entire pre-SN star collapsed to form the black hole. We specify remnant masses from this scenario as M$_{\rm BH,all}$, with ejection fractions of $\fe=0$.

The gaps between these branches are due to the discrete sampling of the S16 simulations. For Branch I, we separately fit the outcomes from core-only implosions and the implosions in which the entire star collapses with simple functional forms. For Branch II, we consider only the outcomes of core-only implosions. We select the functions from a set of power-law or first-, second-, or third-order polynomials, by minimizing the RMS of the residuals. If the RMS of the residuals is $<1\%$ for more than one of the polynomials, we take the lowest-order function. However, we note that the conclusions we find are largely independent of the particular models chosen.

The functions for  M$_{\rm BH,core}$ and  M$_{\rm BH,all}$ differ quite significantly from each other in Branch I because M$_{\rm BH,core}$ depends on the assumed input physics in the stellar modeling but does not depend on the mass loss for this range of progenitor masses, whereas M$_{\rm BH,all}$ is highly sensitive to the particular mass loss prescription and its efficiency. In contrast, for the stars in Branch II, the final masses of the resulting CO-cores are strongly dependent on both the assumed red supergiant and Wolf-Rayet mass loss prescriptions, but the resulting core uniquely determines the black hole mass.

\begin{deluxetable*}{ccccccc}
\tabletypesize{\footnotesize}
\tablewidth{0.8\textwidth}
\tablecaption{Fraction of outcomes that yield black holes in each branch of Fig.~\ref{fig:BHrem} for the five central engine models included in our analysis. }
\tablehead{
\colhead{Branch} & 
\colhead{$X_{\rm BH,W15}$ }  & 
\colhead{$X_{\rm BH,W18}$ }  & 
\colhead{$X_{\rm BH,W20}$ }  &
\colhead{$X_{\rm BH,N20}$ }  &
\colhead{$X_{\rm BH,S19.8}$ }  &
\colhead{$X_{\rm BH,Avg}$ }  
} 
\startdata
I & 0.686 & 0.635  & 0.878 & 0.500 & 0.474 & 0.635 \\
II & 0.875 & 0.750  & 1.00 & 0.500 & 0.500  & 0.725 
\enddata
  \label{table:BHratios}
\end{deluxetable*}

In Branch I, we find that the outcomes from implosions in which only the He-core collapses ($\fe=1$) are well-fit by a linear model and find
\begin{multline}
M_{\rm BH,core}(\Mz) = -2.024 + 0.4130 \Mz, \\
15 \le \Mz \le 40 \Ms.
\end{multline}
with residuals of 0.9\%.

For the outcomes in Branch I for the implosions in which the entire star collapses ($\fe=0$), we find that a third-order polynomial with parameters
\begin{multline}
M_{\rm BH,all}(\Mz)  = 16.28 + 0.00694 (\Mz-21.872) \\
  -0.05973 (\Mz-21.872)^2 + 0.003112 (\Mz-21.872)^3, \\
15 \le \Mz \le 40 \Ms
\end{multline}
is necessary to keep the residuals $\sim$1\%. A second-order polynomial fit produces larger residuals of $\sim$4\%.

In Branch II, we only consider the implosion of the CO-core, since there is no remaining envelope at these high masses. We find that fitting these outcomes with a power-law model is sufficient and find
\begin{multline}
M_{\rm BH,core}(\Mz) = 5.795 + 1.007\times10^{9} (\Mz)^{-4.926}, \\
45 \le \Mz \le 120 \Ms.
\end{multline}
with residuals of $\sim$9\%. This larger residual is due to a single data point. Fitting instead with a second- or third-order polynomial improved the residuals by less than $\sim$0.5\% and made no significant change to the final distribution, so we chose to use the simpler power-law model.

In order to probe regimes in which there may be partial ejection fractions, we can extrapolate from our fits of the special cases of $\fe=0$~or~1, using the approximation
\begin{multline}
\label{eq:Mb}
\Mb(\Mz; \fe) = \fe M_{\rm BH,core}(\Mz) + \\ (1 - \fe)M_{\rm BH,all}(\Mz).
\end{multline}

We use this formalism for its simplicity but also note that the pre-SN core structures and the binding energy outside the He-cores are not identical in models where the stars retain some of their envelopes. As a result, the ultimate fraction of the envelope that gets ejected upon implosion is likely not the same for all progenitor masses.  Considering the uncertain nature of this mechanism, in this work we adopt a simple scenario where $\fe$ has the same value for all applicable stars, in order to explore its effect on the resulting black hole mass distribution.

Using this combined model for each branch, we calculate the probability of a particular black hole remnant mass as
\begin{multline}
\label{eq:MBH}
P(\Mb | \Mz; \fe) =  \\ \left| \frac{d \Mb(\Mz; \fe)}{d \Mz} \right| ^{-1} \delta[\Mz - \Mz(\Mb)],
\end{multline}
where the $\delta$-function encapsulates the relationship between $\Mz$ and $\Mb$ at the value at which the probability distribution is evaluated.
Finally, to get the distribution of black holes for each branch, we marginalize over the progenitor masses, i.e.,
\begin{multline}
\label{eq:integral}
P(\Mb; \fe) = \\ \int P(\Mb | \Mz ; \fe) P(\Mz)  d\Mz,
\end{multline}
where for the probability of finding a particular mass, $\Mz$, in the initial mass distribution, we use the Salpeter IMF, 
\begin{equation}
\label{eq:IMF}
P(\Mz) = C (\Mz)^{\alpha},
\end{equation}
with $\alpha$=-2.3 and $C$=0.065 \citep{Salpeter1955}.

We weight the probability of each branch by a value, $X_{\rm BH}$, which represents the fraction of outcomes in that branch that produced black holes. These weighting fractions, which are shown in Table~\ref{table:BHratios} for each engine, reflect the fact that the explosion outcomes can form either neutron stars or black holes in certain mass ranges, as discussed in $\S$\ref{sec:models}, and need to be treated probabilistically. Note that we do not include fallback cases in the number of successful black holes, but do include them in the number of possible outcomes.

\subsection{Comparing the simulated and observed black hole mass distributions}
\begin{figure}[ht]
\centering
\includegraphics[width=0.45\textwidth]{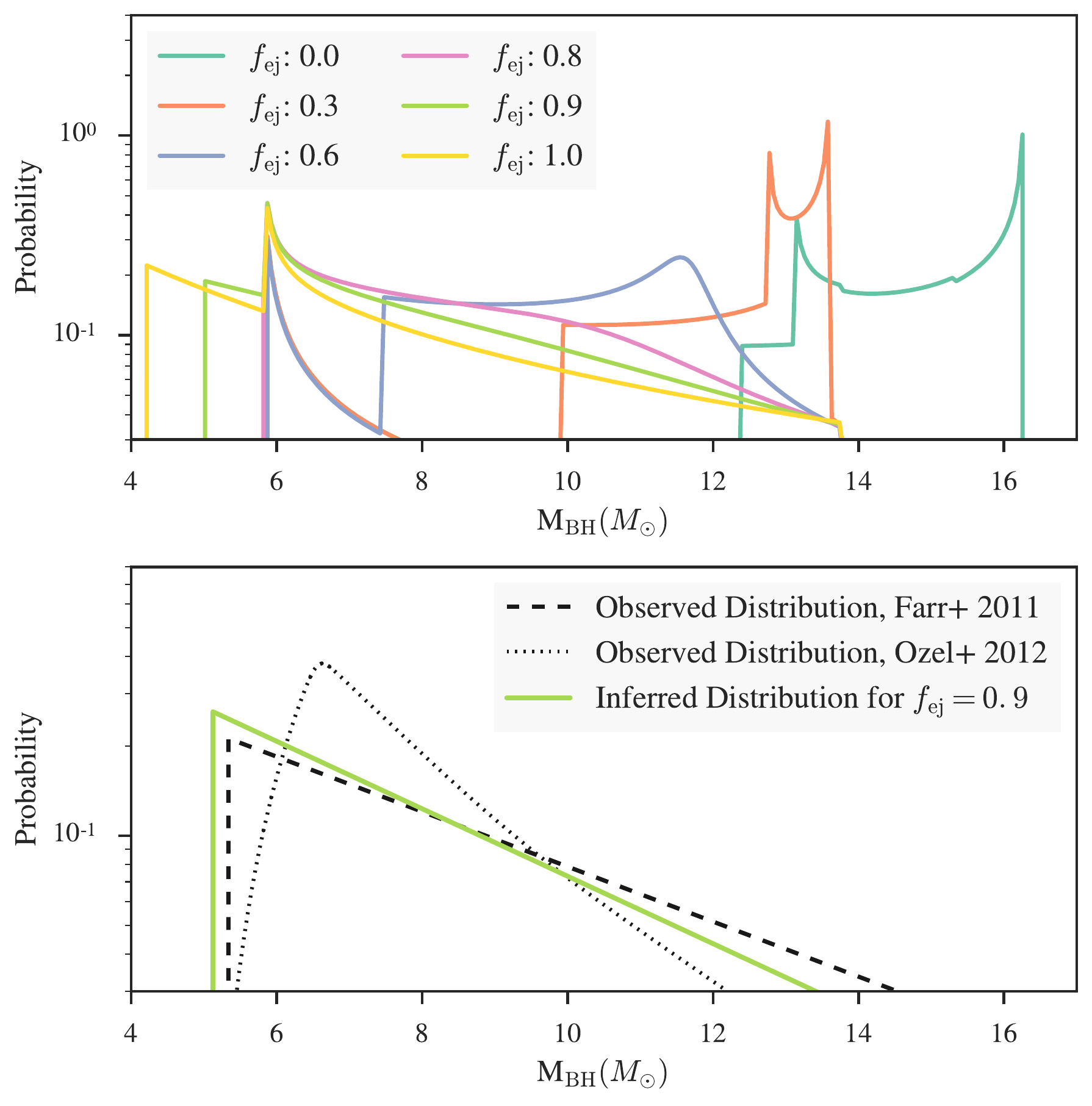}
\caption{\label{fig:PBH} \textit{Top: }Mass distribution of black holes averaged from the simulations for all five central engines. Different colors represent different fractions of the stellar envelope that are ejected either during the implosion, by a weak shock, or prior to the implosion, via tidal stripping by a binary companion. \textit{Bottom: } Comparison to the observed mass distributions of \citet{Farr2011} and \citet{Ozel2012}, shown in the black dashed and dotted lines, respectively. Here, the green line represents the distribution that would be inferred from the underlying simulated distribution for $\fe=0.9$, if a decaying exponential form is assumed. We find that to recreate the observed distribution, a relatively large ejection fraction is required.}
\end{figure}

We calculate the final black hole mass distribution by summing the probability contributions for each branch, as found in eq.~(\ref{eq:integral}), and weighting each contribution by the ratios, $X_{\rm BH}$, of Table~\ref{table:BHratios}. We include in this analysis the results from each of the five central engines, which we average together. We show the resulting black hole mass distribution, for various ejection fractions, in Figure~\ref{fig:PBH}.

We find that, in general, the smaller the ejection fraction, the narrower the expected mass distribution. This is because the pre-SN mass is less sensitive to the initial mass in Branch I than the He-core mass is, as shown in Fig.~\ref{fig:BHrem}. Additionally, we find that smaller ejection fractions produce larger black holes, as expected.  For an ejection fraction of 0, the distribution is confined to $\Mb\sim12-16~\Ms$, with sharp peaks at $\sim 13$ and 16~$\Ms$. In contrast, we find that an ejection fraction of 0.9 leads to a mass distribution with a soft decay, spanning from $\Mb \sim5-12~\Ms$. 

In the bottom panel of Fig.~\ref{fig:PBH}, we show two black hole mass distributions inferred from the observed black hole masses in our Galaxy.
The first distribution was calculated in \citet{Ozel2010b} from black hole masses measured from 16 low-mass X-ray binaries. The resulting distribution was well-fit by a decaying exponential. The second distribution was calculated similarly by \citet{Farr2011} from mass measurements of 15 black holes in low-mass X-ray binaries and 5 black holes in high-mass, wind-fed X-ray binaries. In this study, they fit several different models to the data and found strong evidence for an exponentially decaying profile of the form,  
\begin{multline}
P(\Mb; M_{\rm min}, M_0) = \frac{\exp(M_{\rm min} / M_0)}{M_0} \\ \times
   \begin{cases} 
	 \exp(-\Mb/ M_0), & \Mb \ge M_{\rm min}  \\
	 0, & \Mb < M_{\rm min}  
   \end{cases},
\label{eq:Mfarr}
\end{multline}
where $M_{\rm min}$ is the minimum black hole mass, which was found to be 5.3268~$\Ms$, and $M_0$ is a scale mass found to be 4.70034~$\Ms$ \citep{Farr2011}.

In order to make the comparison between these observed distributions and our simulated distribution more directly, we note that the substructure in the simulated distributions is finer than the typical uncertainties in the observations and would not be observable as is. Thus, we also calculate the distribution that would be inferred from the underlying simulated distribution, by drawing 200 random  black hole masses from the simulated distribution and fitting them with the exponential decay model of eq.~(\ref{eq:Mfarr}). This ``inferred," simulated distribution is shown for $\fe=0.9$ in the bottom panel of Fig.~\ref{fig:PBH}, and shows close agreement with the observed distribution in our Galaxy.

\begin{figure}[ht]
\centering
\includegraphics[width=0.45\textwidth]{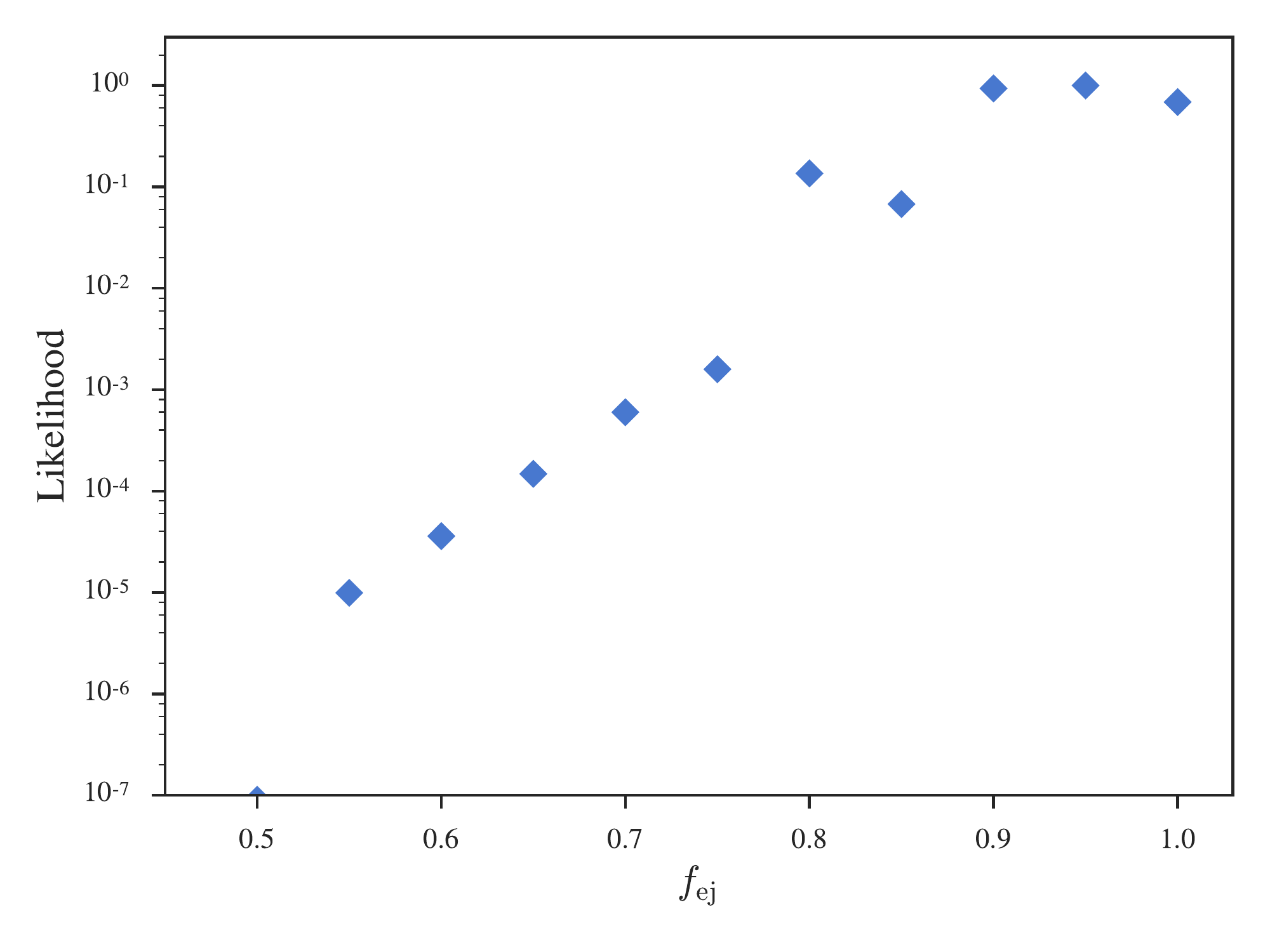}
\caption{\label{fig:resids} Likelihood that the inferred, simulated distribution matches the observed distribution of black hole masses, for various ejection fractions. We find that the likelihood peaks when a relatively large fraction of the stellar envelope has been ejected, $\fe~>~0.9$.}
\end{figure}

Finally, we calculate the likelihood of the ``inferred" distributions for each ejection fraction, assuming the same functional form of eq.~(\ref{eq:Mfarr}). We calculate the likelihood as
\begin{equation}
\mathcal{L} = \exp \left[ - \sum_i \frac{(P_{\rm inferred}(M_{\rm BH,i}; \fe) - P_{\rm obs}(M_{\rm BH,i}))^2}{P_{\rm obs}({M_{\rm BH,i}})} \right],
\label{eq:likelihoods}
\end{equation}
where $P_{\rm obs}(M_{\rm BH,i})$ is the probability given in eq.~(\ref{eq:Mfarr}) for the inferred parameters from \citet{Farr2011}, for black holes above the minimum mass of \citet{Farr2011} in our sampling. We show these likelihoods in Fig.~\ref{fig:resids}, and find that ejection fractions of $\fe \gtrsim 0.9$ have the highest likelihoods, given the observed mass distribution. This implies not only that the S16 simulations are closely reproducing the black hole masses observed in our Galaxy, but also that a large fraction of the stellar envelope must be ejected in order to form the observed black holes, either during the implosion itself or beforehand, through tidal stripping from a binary companion.

Finally, we note that, for all ejection fractions in the solar-metallicity models of S16, there appear to be no black holes with masses above $12-16~\Ms$. This is particularly interesting in light of the recent inferences of black holes with $M\ge22~\Ms$ that have been made with the first LIGO and Virgo gravitational wave detections \citep{Abbott2016b, Abbott2016, Abbott2017, LIGOVirgo2017}, as can be seen in Fig.~\ref{fig:graveyard}. We discuss this further in $\S$\ref{sec:LIGO}.

\section{Neutron star mass distribution}
\label{sec:NSdist}
\begin{figure*}[ht]
\centering
\includegraphics[width =0.9 \textwidth]{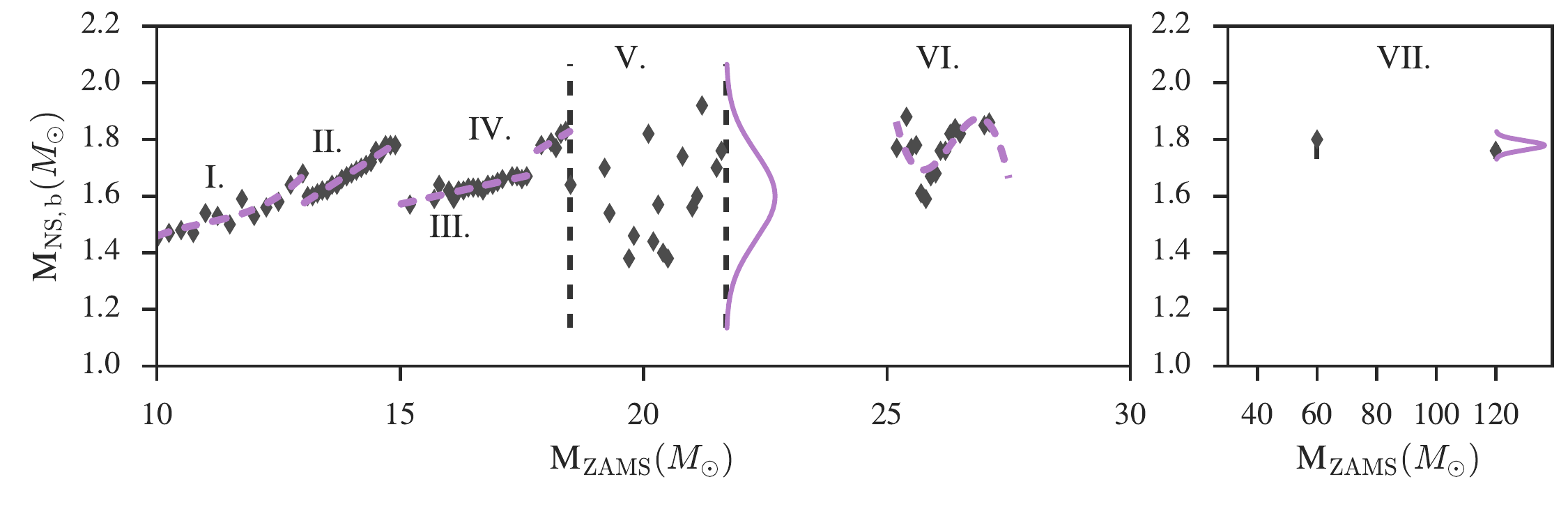}
\caption{\label{fig:NSrem} Neutron star baryonic masses as a function of the progenitor mass, for engine W18. We identify 7 distinct branches in this distribution. The analytic functions that we fit to each branch are shown in purple.}
\end{figure*}

We calculate the neutron star mass distribution with the same method that we used for the black hole distribution of $\S$\ref{sec:BHdist}. Figure~\ref{fig:NSrem} shows the neutron star remnant masses for various progenitors, as calculated with the W18 engine. The neutron star masses produced by the five different engines that we used in $\S$\ref{sec:BHdist} are offset slightly from one another, although each give approximately similar results. As a result, in order to avoid introducing artificial noise by combining these slightly different sets of outcomes, we only include engine W18 in the following analysis and take it to be representative of all five models.

Within the neutron star remnant masses, we identify 7 distinct segments that we fit with simple analytic functions, as in $\S$\ref{sec:BHdist}. We show the analytic functions that we fit to each branch in Fig.~\ref{fig:NSrem} in purple.

We find that the first branch is best fit by a third-order polynomial with parameters
\begin{multline}
\Mnb(\Mz) = 2.24 + 0.508 (\Mz-14.75) \\ + 0.125 (\Mz-14.75)^2 + 0.0110(\Mz-14.75)^3, \\
9 \le \Mz \le 13 \Ms.
\end{multline} 
The RMS of the residuals to this fit is $\sim1.3\%$. Here, $\Mnb$ is the baryonic mass of the neutron star. The baryonic masses are the natural output of the stellar evolution and explosion models, which we will later convert to gravitational masses. 

We find that Branches~II$-$IV are sufficiently fit with linear models, with residuals $<1\%$ for
\begin{multline}
\Mnb(\Mz) = 0.123 + 0.112 \Mz\\
13 < \Mz < 15 \Ms,
\end{multline}
\begin{multline}
\Mnb(\Mz) = 0.996 + 0.0384 \Mz \\
15 \le \Mz < 17.8 \Ms.
\end{multline}
and
\begin{multline}
\Mnb(\Mz) = -0.020 + 0.10 \Mz \\
17.8 < \Mz  <18.5 \Ms.
\end{multline}

We find Branch~V to be approximately randomly distributed, and thus fit it with a Gaussian distribution, i.e.,
\begin{equation}
\label{eq:NSgauss}
P(\Mnb | \Mz) = \frac{1}{\sqrt{2 \pi} \sigma} \exp{[-(\Mnb - M_0)^2 / 2 \sigma^2]},
\end{equation}
where $M_0$ and $\sigma$ are the mean and standard deviation of the distribution.
For Branch~V ($18.5 \le \Mz  < 21.7 \Ms$), we find the standard deviation to be $\sigma = 0.155$ and the mean to be $M_0=1.60~\Ms$.

Branch VI is best fit by a third-order polynomial with parameters 
\begin{multline}
\Mnb(\Mz) = 3232.29 - 409.429(\Mz - 2.619) \\ 
	+ 17.2867 (\Mz-2.619)^2 - 0.24315(\Mz-2.619)^3, \\
25.2 \le \Mz < 27.5 \Ms,
\end{multline}
with residuals $\sim$3\%. 

Finally, we fit Branch~VII ($60 \le \Mz \le 120 \Ms$) with a Gaussian distribution and find $\sigma = 0.016$ and $M_0 = 1.78~\Ms$. It should be noted, however, that Branch~VII contains only two points; as a result, the parameters of this particular fit should be interpreted with caution. We show the Gaussian distributions on the right side of each $\Mz$ range in Fig.~\ref{fig:NSrem}.

We use the analytic functions for Branches~I$-$IV and VI to calculate the probability distribution of neutron star masses, according to
\begin{multline}
\label{eq:MNS}
P(\Mnb | \Mz) = \\  \left| \frac{d \Mnb(\Mz)}{d \Mz} \right| ^{-1} \delta[\Mz - \Mz(\Mnb)], 
\end{multline}
For Branches V and VII, we simply use the fitted Gaussian distribution for $P(\Mnb | \Mz)$.

\begin{figure}[ht]
\centering
\includegraphics[width=0.5\textwidth]{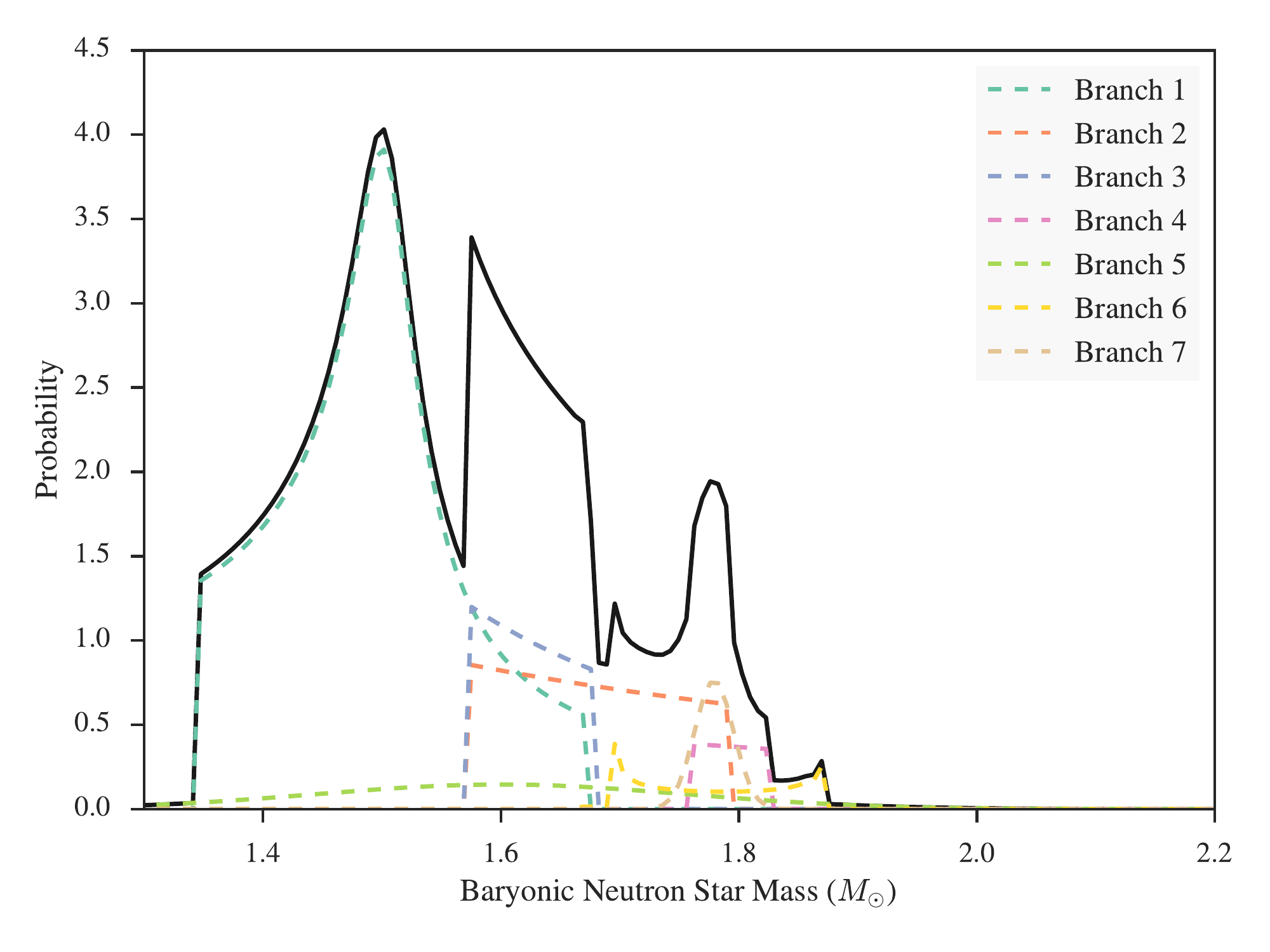}
\caption{\label{fig:PNSbar} Baryonic mass distribution of neutron stars from the S16 simulations. The overall distribution is shown in black. The various dashed colors represent the contributions from each branch of ZAMS progenitors.}
\end{figure}

Finally, we marginalize the probabilities of each branch by
\begin{equation}
\label{eq:integralNS}
P(\Mnb) = \int P(\Mnb | \Mz) P(\Mz) d\Mz,
\end{equation}
and use the IMF of eq.~(\ref{eq:IMF}) for $P(\Mz)$, as in $\S$\ref{sec:BHdist}. We calculate the total distribution by summing the probability contributions from each branch, weighted by the fraction of outcomes that produce neutron stars ($X_{\rm NS}$) in that branch. We list these fractions in Table~\ref{table:NSratios}. We show the neutron star baryonic mass distribution, along with the contributions from each branch, in Fig.~\ref{fig:PNSbar}.

\begin{deluxetable}{cc}
\tabletypesize{\footnotesize}
\tablewidth{0.3\textwidth}
\tablecaption{Fraction of outcomes that yield neutron stars in each branch of Fig.~\ref{fig:NSrem}. }
\tablehead{
\colhead{Branch} &
\colhead{$X_{\rm NS,~W18} $ } 
}
\startdata
    I &   1.00 \\ 
    II &   1.00 \\
    III &  0.679 \\
    IV & 0.833 \\
    V &  0.500 \\ 
    VI &  0.652 \\ 
    VII &  0.400  
\enddata
  \label{table:NSratios}
\tablecomments{Only the W18 engine results were included in our analysis of neutron star distributions.}
\end{deluxetable}

We convert from the baryonic ($\Mnb$) to the gravitational ($\Mng$) mass distribution with the transformation
\begin{equation}
P(\Mng) = P(\Mnb) \left| \frac{d \Mnb}{d \Mng} \right| ^{-1},
\end{equation}
where $\Mng$ is the gravitational mass and we calculate the derivative using the relationship between binding energy (BE) and baryonic and gravitational masses,
\begin{equation}
M_{b} = M_{g} + \rm BE.
\end{equation}
For the binding energy, we use the \citet{Lattimer2001} approximation
\begin{equation}
BE = \Mng \times \left( \frac{0.6 \beta}{1- 0.5\beta} \right),
\end{equation}
where $\beta \equiv G M_G / R c^2$ is the neutron star compactness. We find that the gravitational mass distribution depends only weakly on the choice of radius in the binding energy approximation, so we use a characteristic value of 12~km. 

\subsection{Comparing the simulated and observed pulsar mass distributions}
\begin{figure}[ht]
\centering
\includegraphics[width=0.5\textwidth]{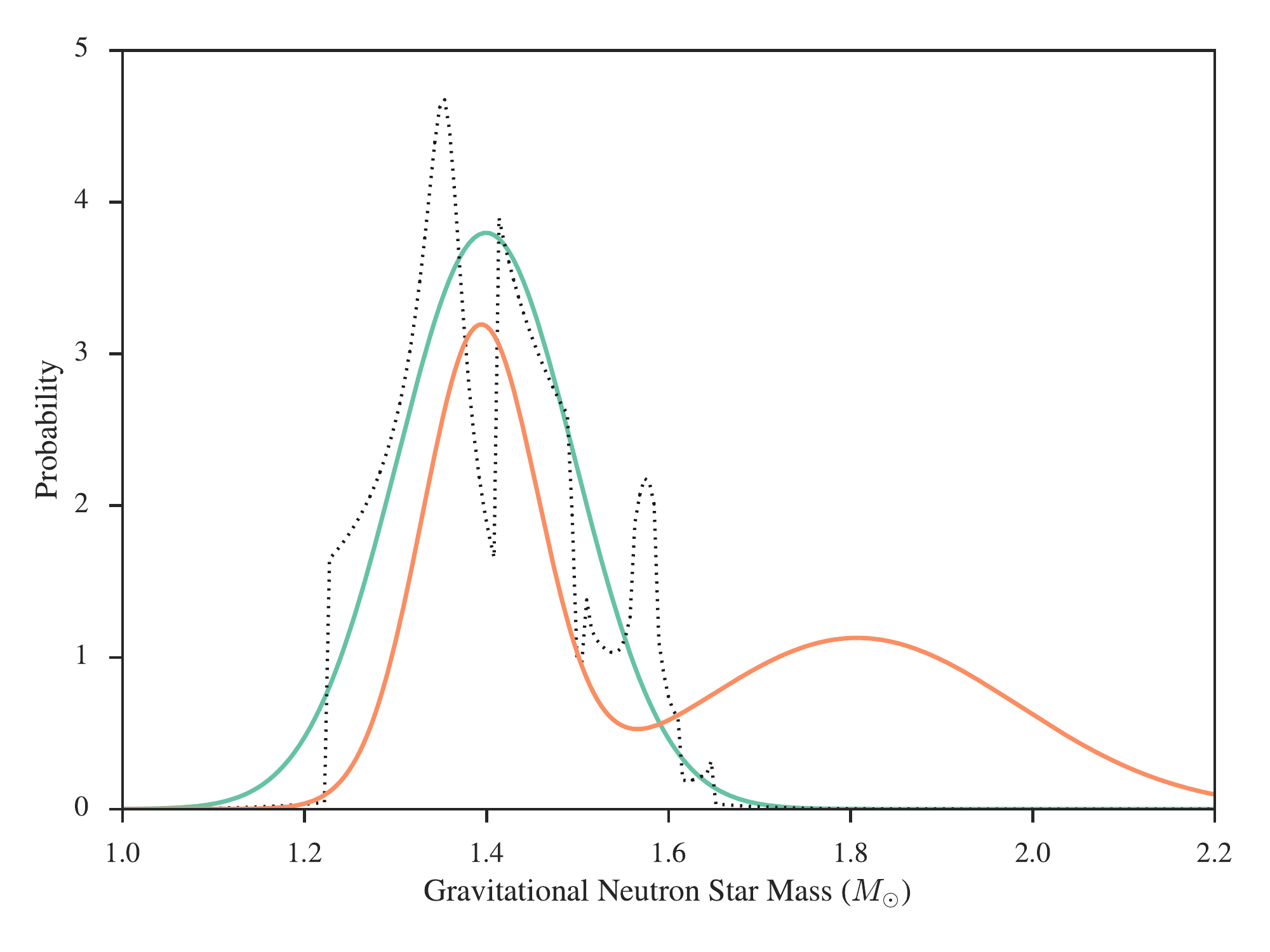}
\caption{\label{fig:PNSgrav} The gravitational mass distribution of neutron stars predicted from the S16 simulations (black dotted line) and the Gaussian distribution that would be inferred from the mock data that we produced from the full simulated distribution (green line). The orange line shows the observationally determined mass distribution of millisecond pulsars from \citet{Antoniadis2016}. We find that the simulated mass distribution aligns very closely with the low-mass component of the observed distribution. }
\end{figure}

We calculate the gravitational mass distribution as described in $\S$\ref{sec:NSdist} and show the outcome as the dotted line in Fig.~\ref{fig:PNSgrav}. However, because the substructure between the various peaks is finer than could be detected with realistic observational uncertainties as before, we also compute and show in the same figure the Gaussian distribution that would be inferred from this underlying distribution. We calculate this Gaussian by drawing 200 points from the underlying distribution and fitting with a single Gaussian function, i.e.,
\begin{equation}
P(\Mng) = \frac{1}{\sqrt{2\pi}\sigma}\exp^{-(M-M_0)/2\sigma^2}.
\end{equation}
We show the most-likely value for the mean, $M_0$, and standard deviation, $\sigma$ in Table~\ref{table:grav} and the resulting distribution in green in Fig.~\ref{fig:PNSgrav}. Finally, we also include in Fig.~\ref{fig:PNSgrav} the observed neutron star mass distribution inferred from 33 millisecond pulsars in \citet{Antoniadis2016} in orange. The parameters of the low-mass peak of the \citet{Antoniadis2016} distribution, shown in Table~\ref{table:grav}, agree within 1-$\sigma$ with the inferred parameters from the S16 distribution.

We can determine the origin of the low-mass peak of the observed distribution by comparing with the simulated distributions of each branch, which are shown in Fig.~\ref{fig:PNSbar}. We find that progenitors in Branches I$-$III, i.e. with $\Mz = 9-17.8 \Ms$, are the dominant contributors to the low-mass peak that agrees well with the observed one. 

We also see that the narrow peak at $\Mng\sim~1.6~\Ms$ originates primarily from progenitors in Branch~VII, with a modest enhancement from Branch IV. Because branch~VII contains only two neutron stars that are fit with a Gaussian, its properties could easily be affected by a larger number of simulations in that mass range and it should be interpreted with caution. Finally, Branches V and VI contain very broad distributions and correspondingly do not contribute significantly to any particular peak.

In the overall comparison of neutron star masses, we find excellent agreement between the simulated distribution and the low-mass peak of the observed mass distribution. We wish to emphasize here that this is a highly constrained comparison, with no parameters that can be adjusted in either distribution to improve their relative alignment. As described in $\S$\ref{sec:models}, the simulations are calibrated only to reproduce the energetic properties of SN~1987A and SN~1054. The alignment that we see in Fig.~\ref{fig:PNSgrav}, in addition to the alignment in the black hole mass distribution, therefore, seem to be a natural consequence of calibrating to the global energetic properties. Therefore, we argue that this level of agreement in the mass distributions provides a strong and independent validation of these stellar evolution and explosion models.

\begin{deluxetable}{ccc}
\tabletypesize{\footnotesize}
\tablewidth{0.4\textwidth}
\tablecaption{Gaussian Parameters for low-mass neutron star peak in Fig.~\ref{fig:PNSgrav} }
\tablehead{
\colhead{ Source of distribution } &
\colhead{ $M_0$ ($\Ms$) } &
\colhead{ $\sigma$} 
}
\startdata
    \citet{Antoniadis2016} observations &   1.393 & 0.064  \\ 
    S16 simulations &   1.399  & 0.098
\enddata
  \label{table:grav}
\end{deluxetable}

\section{Missing high-mass remnants}
\label{sec:highM}
In the analyses of $\S$\ref{sec:BHdist} and \ref{sec:NSdist}, we found that the remnant mass distributions predicted by the S16 simulations show close agreement with both the observed black hole mass distribution and low-mass neutron star distribution, offering new evidence in support of these models. However, these simulations do not produce the high-mass LIGO black holes that can be seen in Fig.~\ref{fig:graveyard} and are missing the high-mass component of the observed pulsar mass distribution. In this section, we explore possible explanations for these discrepancies.

\subsection{High-mass black holes}
\label{sec:LIGO}
The recent detection of gravitational waves from black hole binary GW150914 provided the first observational evidence of ``heavy" stellar-mass black holes ($M \gtrsim 25~\Ms$). The black holes in this binary were inferred to have masses of 29$^{+4}_{-4}$ and 36$^{+5}_{-4}$~$\Ms$ \citep{Abbott2016}. In an initial characterization, \citet{Abbott2016a} proposed that the formation of such massive black holes via single-star evolution requires weak winds, which in turn requires an environment with metallicity $Z~\lesssim~1/2~Z_{\odot}$. Subsequent detections have found further evidence of additional ``heavy" stellar-mass black holes \citep{Abbott2016b, Abbott2017, LIGOVirgo2017}.

\begin{figure}[ht]
\centering
\includegraphics[width=0.48\textwidth]{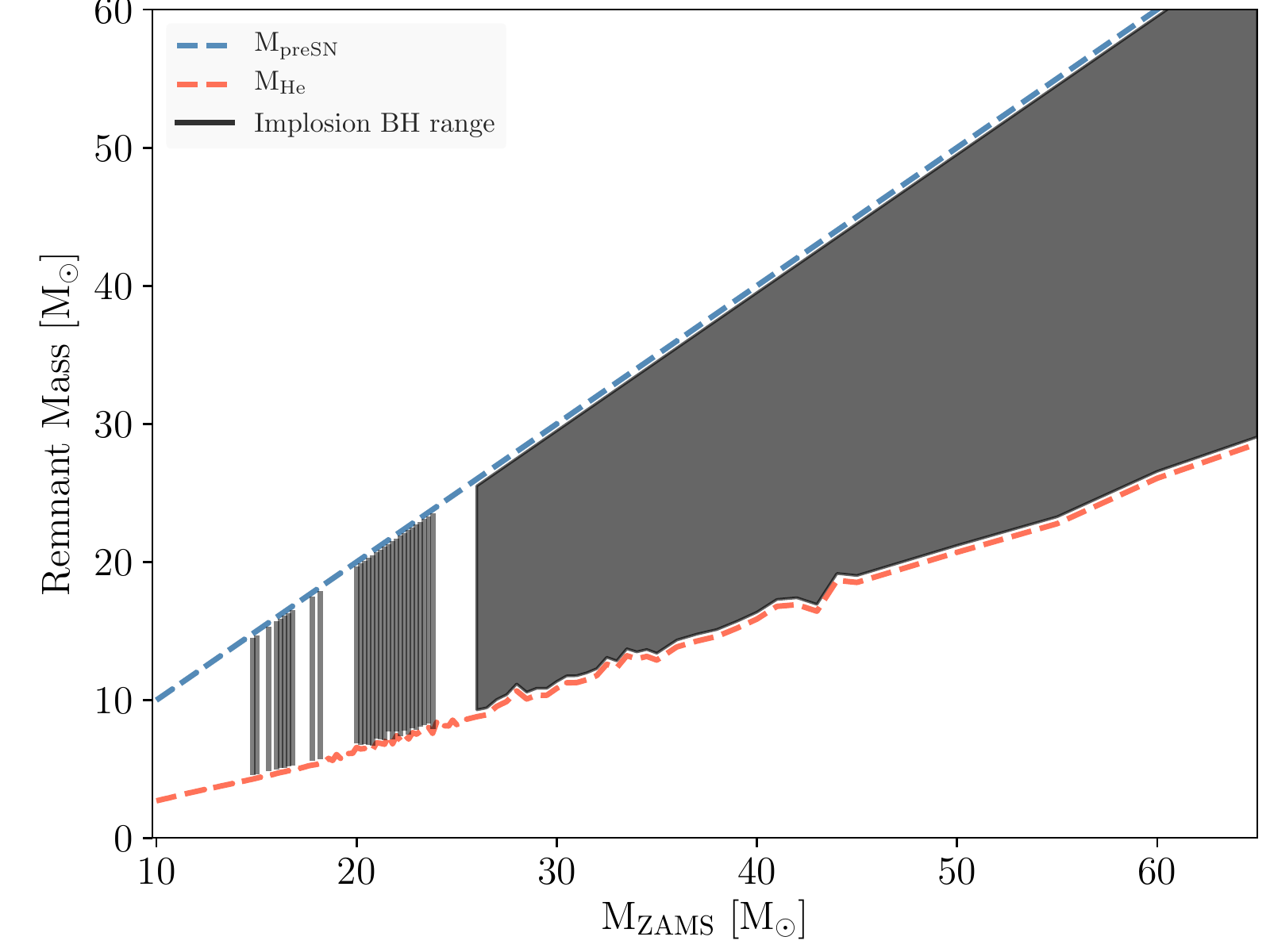}
\caption{\label{fig:lowZ} The range of possible black hole masses, bound by the pre-SN and He core masses of imploding progenitors, are shown for the ultra-low $10^{-4}~Z_\odot$ metallicity models and a sample central engine (N20). The input physics in these models are identical to the those employed in this paper, with the exception of the reduced metallicity. Compared to the models with solar metallicity shown in Fig.~\ref{fig:BHrem}, the mass loss here is negligible and thus the implosions from the heaviest models allow the formation of heavy stellar mass black holes.}
\end{figure}

The mass range of implosion outcomes due to a sample central engine applied to low metallicity progenitors is illustrated in Figure~\ref{fig:lowZ}. The models are the ultra-low metallicity ($10^{-4}Z_\odot$) ``U-series'' set from \citet{Sukhbold2014}, which consists of 110 models with initial masses between 10 and 65~$M_\odot$. At such a low metallicity, mass loss is negligible and both the He core and final pre-supernova masses increase monotonically with initial mass. At lower initial mass, the explosion landscape is similar to the solar metallicity models, since the pre-supernova core structure in these stars are not strongly affected by metallicity. For more massive models, however, the cores are significantly harder to explode. Indeed, with the adopted sample central engine, all stars implode above $\Mz>30\ M_\odot$. From this figure, it is clear that black holes can be formed with $\Mb \gtrsim 25\Ms$, when the metallicity is sufficiently reduced.

An alternate channel for heavy stellar-mass black hole formation from single stars has recently been proposed in the form of strong magnetic fields. \citet{Petit2017} showed that for progenitor stars with $\Mz$ in the range $40-80~\Ms$, the presence of strong surface magnetic fields can significantly quench mass loss by magnetically confining a fraction of wind material to the star's surface. For a strongly magnetic 80~$\Ms$ star, this reduces the mass lost during the main-sequence evolution by 20~$\Ms$, which is the equivalent mass loss reduction achieved by reducing the stellar metallicity to $Z\sim1/30~Z_{\odot}$ for a non-magnetic star \citep{Petit2017}. The S16 simulations do not include magnetic fields or rotation, but such a model offers another possible mechanism for producing black holes in the regime that was probed by the LIGO detection, without reducing the metallicity.

Numerous studies have also explored the effect of binary evolution for producing heavy stellar mass black holes. Typically, simulations of binary massive star evolution result in the formation of a common envelope, via Roche lobe overflow (e.g., \citealt{Voss2003}). Accretion through the common envelope could, in principle, increase the mass of a star's He core and hence the mass of the post-SN black hole remnant to what was observed in the LIGO detections \citep{Belczynski2016, Kruckow2016, Woosley2016, Eldridge2016, Stevenson2017}. However, there remain many uncertainties in the common-envelope physics used that affect the possible outcomes (see, e.g., \citealt{Ivanova2013}).

It is possible to avoid the uncertainties and pitfalls of the common envelope scenario by requiring a close binary orbit. In this mechanism, the close companions tidally spin up one another. The rapid rotation then induces mixing that is faster than the chemical gradient build-up due to nuclear burning, so that the stars remain chemically homogeneous throughout hydrogen burning. This keeps the stars from developing massive hydrogen envelopes and thus offers a way in which the stars could evolve to black holes without ever undergoing significant mass transfer. It has been shown that in such models, it is indeed possible to form heavy black holes, with $M~\gtrsim~25~\Ms$  \citep{de-Mink2016, Mandel2016, Marchant2016}. However, there remain large uncertainties in the efficiency of the mixing processes involved and in the impact of stellar winds on the orbital evolution, making it unclear whether this channel is likely or even possible \citep{Mandel2016}. 

Finally, many studies have found that dynamical assembly of black hole binaries in dense stellar clusters can also produce more massive black holes, via multi-body encounters, mass segregation, and gravitational focusing. These processes favor heavier black holes, which are already easier to form in the low-metallicity environments of globular clusters \citep{Mapelli2016,OLeary2016,Rodriguez2016, Askar2017, Park2017}. If this is the primary way in which LIGO-mass black holes are formed, the single-star evolution framework of S16 and this paper would not apply.

As a final remark, there may still be an upper bound on the expected masses of ``heavy" stellar-mass black holes that might play a role in the above formation mechanisms. A recent analysis by \citet{Woosley2017} predicts that no black holes with masses between $\sim$52 and 133 $\Ms$ should be found in nature in close binary systems due to pulsational pair-instability effects. While this is in agreement with the current massive black hole detections by the LIGO-Virgo collaboration, mergers within a Hubble time from more complicated systems with more than two components could violate this bound.

\subsection{Missing high-mass neutron stars}
In $\S$\ref{sec:NSdist}, we also found that the high-mass peak of the observed neutron star mass distribution of \citet{Antoniadis2016} was not reproduced by the S16 simulations, despite very close agreement in the low-mass regime.
Indeed, the S16 simulations do not produce any neutron stars with masses above $\Mng > 1.7\ \Ms$ and it is possible that the lack of high-mass neutron stars could imply incomplete physics in the stellar evolution models. Recent work by \citet{Sukhbold2017}, which employs updated physics and a denser grid of models, finds the pre-SN core structures to be intrinsically multi-valued, including for the mass range $14<\Mz<19\ \Ms$. Without performing full explosions of the pre-SN models, that study finds tantalizing evidence that neutron stars with $\Mng\sim1.9\ \Ms$ can be made by the most massive stars with a significant second oxygen shell burning ($\Mz\sim15\ \Ms$). Whether these new models can recreate the high-mass peak of neutron star distribution will be explored in a future work. 

Alternatively, the discrepancy would disappear if the observed high-mass peak is due to accretion rather than a second population of neutron star birth masses. However, \citet{Antoniadis2016} argue against such a path: they highlight several examples of high-mass neutron stars with companions that would be too small to allow significant accretion, inferring that the birth masses must be $\gtrsim 1.7~\Ms$. Even if we allow some accretion and lower the high-mass component to $1.7~\Ms$, such a population of neutron stars is still missing from the simulations. We discuss two other possibilities below.

\subsubsection{Effect of rotation}
\label{sec:rot}
More massive neutron stars are generally made by more massive main sequence stars, yet only a few models with $\Mz>30\ \Ms$ successfully explode in the neutrino-driven formalism of S16. The effects of rotation are expected to be important in the deaths of these heavier stars \citep[e.g.,][]{Heger2005}, and therefore the inclusion of rotation in the modeling of both the evolution and explosion may result in more successful explosions at higher initial mass and consequently in more massive neutron stars.

\subsection{Effect of binary evolution}
\label{sec:binary}
It is also possible that the explodability of the pre-SN stars and the resulting compact object masses are influenced by binary evolution. The potential impact of binary evolution is particularly important to consider in our comparison, since the observed black hole and pulsar masses all come from binary systems, while the models of S16 assume single-star evolution. 

While the reproduction of observed compact object mass distributions in this study may suggest that binary effects are negligible, such an argument is not conclusive. As an example, single-star models had historically reproduced the observed populations of massive stars, even though binary effects were known to be important in $\sim$70\% of those stars. It was eventually shown that the assumed mass loss rates had been set 3$-$10 times too high in the single-star models and were effectively compensating for mass loss due to binary Roche-lobe overflow or common envelope evolution (see \citealt{Smith2014} for a review). It is natural to ask whether the single-star models of S16 may similarly include physics that is mimicking binary effects.

Given the uncertain nature of mass loss \citep[e.g.,][]{Renzo2017}, the prescriptions employed in the models of S16 may well be overestimating what is really experienced by single massive stars. In the current work, we cannot quantify to what extent the final pre-SN masses of the S16 progenitors or the ejections fractions inferred in $\S$\ref{sec:BHdist} are influenced by binary effects. However, we can qualitatively say that the net mass loss effects from binaries would have to closely match the combined result from the employed mass loss prescriptions \citep{Nieuwenhuijzen1990,Wellstein1999} and the relatively high envelope ejection fraction of $\fe\sim$0.9, in order to reproduce the agreement we find with the observed remnant masses. 

Another important consideration is the assumed initial mass function. The formation of binary systems requires a specific set of conditions, and the formation of binaries that remain bound even after one member explodes requires an even more restrictive scenario. The mass function of binaries that can produce compact objects with bound companions, therefore, might differ from that of isolated stars. Throughout our analysis, we used the Salpeter IMF. It is possible that using a more representative mass function could increase the weighting given to the high-mass stars, and perhaps enhance a high-mass peak of neutron stars. Fully exploring such a binary mass function could be carried out using detailed population synthesis models. We note, however, that in the case of the high-mass neutron star peak, given the absence of neutron stars with masses above 1.7~$\Ms$ in the S16 results, modifying the initial mass function alone is unlikely to produce the missing high-mass peak.

\section{Conclusions}
In this paper, we directly confronted the outcomes from the stellar evolution models and neutrino-driven explosion simulations of S16 with the observed neutron star and black hole mass distributions. Given that the central engines of the simulations were calibrated only to reproduce the $^{56}$Ni mass, kinetic energy, and neutrino burst timescale of SN~1987A and the kinetic energy of SN~1054, it is perhaps surprising that the remnant mass distribution from these simulations agrees so closely with the observed black hole mass distribution and the low-mass distribution neutron stars. This degree of agreement can be taken as evidence that the stellar evolution and explosion models we studied here have reached a point where they are accurately capturing the relevant stellar, nuclear, and explosion physics involved in the formation of compact objects.

In comparing the simulated and observed mass distributions, we find that the stellar evolution and explosion models are able to accurately reproduce the observed black hole distribution \citep{Ozel2010b, Farr2011}, if a large fraction of the stellar envelope is ejected during the SN ($\fe \sim~0.9$). However, the solar-metallicity models we use in this paper do not produce heavy stellar-mass black holes, the existence of which have recently been confirmed by the LIGO gravitational wave detections. We show that similar models to those used in S16 can indeed produce heavier black holes, if the metallicity is sufficiently reduced. We also review alternate mechanisms that may produce such black holes, including via rapid rotation in binary evolution or strong magnetic fields in single-star evolution, but large uncertainties remain in the current understanding of these mechanisms.

We also find very close agreement between the simulated distribution of neutron star masses and the low-mass peak of the observed bimodal distribution found by \citet{Antoniadis2016}; specifically, that the simulated and observed Gaussian distributions agree to within their 1-$\sigma$ uncertainties. From the S16 simulations, we determine that the low-mass neutron stars originate from progenitors with $\Mz \approx9-18~\Ms$. However, the simulations do not reproduce the observed high-mass peak at $\Mng\sim1.8~\Ms$. In fact, the simulated distribution ends below $\Mng\sim1.7~\Ms$. We explore several possibilities for this discrepancy, including that the high-mass formation channel might require consideration of the binary mass function (as opposed to the single-star IMF we use here), or consideration of additional physics, such as the impact of rotation on the explodability of high-mass progenitors.

The method we have developed here, of directly confronting the simulated outcomes with measured mass distributions, will allow further tests of new models and ultimately will allow us to better understand the formation of compact objects. With the framework developed here, other formation channels may be tested as well, offering a new way to constrain stellar evolution and explosion models.
\\\\{\em{Acknowledgements.\/}} We thank Thomas Ertl for his help with the simulations used in this study and for many discussions on this work. We thank Dimitrios Psaltis, Nathan Smith, and Jeremiah Murphy for helpful discussions and comments on this work. We also thank Casey Lam for their input on our reported fit coefficients. CR is supported by the NSF Graduate Research Fellowship Program Grant DGE-1143953. TS is partially supported by NSF grant PHY-1404311 to John Beacom. FO gratefully acknowledges a fellowship from the John Simon Guggenheim Memorial Foundation in support of this work as well as support by the NASA grant NNX16AC56G.

\bibliography{massbib}
\bibliographystyle{apj}

\end{document}